\documentclass[runningheads]{llncs}
\usepackage[T1]{fontenc}
\usepackage{mdframed}
\usepackage{hyperref}
\usepackage{graphicx}
\begin{document}
\title{Large Language Models Playing Mixed Strategy Nash Equilibrium Games}
%
%
\author{Alonso Silva\inst{1} }
\institute{Nokia Bell Labs, France\\
\email{alonso.silva@nokia-bell-labs.com}\\
\url{https://www.bell-labs.com/about/researcher-profiles/alonsosilva/}
}
\maketitle              
\begin{abstract}
Generative artificial intelligence (Generative AI), and in particular Large Language Models (LLMs), has gained significant popularity
among researchers and industrial communities, paving the way
for the integration of LLMs in different domains, such as robotics,
telecom, and healthcare.
In this paper, we study the intersection of game theory and generative artificial intelligence, focusing on the capabilities of LLMs to find the Nash equilibrium in games with a mixed strategy Nash equilibrium and no pure strategy Nash equilibrium (that we denote {\sl mixed strategy Nash equilibrium games}). 
The study reveals a significant enhancement in the performance of LLMs when they are equipped with the possibility to run code and are provided with a specific prompt to incentivize them to do so. 
However, our research also highlights the limitations of LLMs when the randomization strategy of the game is not easy to deduce. It is evident that while LLMs exhibit remarkable proficiency in well-known standard games, their performance dwindles when faced with slight modifications of the same games. 
This paper aims to contribute to the growing body of knowledge on the intersection of game theory and generative artificial intelligence while providing valuable insights into LLMs strengths and weaknesses. It also underscores the need for further research to overcome the limitations of LLMs, particularly in dealing with even slightly more complex scenarios, to harness their full potential.

\end{abstract}
\section{Introduction}
Generative artificial intelligence (Generative AI) has emerged as a dynamic field within AI, empowering machines with algorithms that enable them to generate unique content, including music, images, code, text, and human-like conversations. One notable example is Large Language Models (LLMs), a specific type of Generative AI model that undergoes training on extensive unlabeled textual datasets. LLMs have demonstrated remarkable abilities in various domains, including question answering, translation, summarization, code generation, etc.~\cite{yenduri2023}.
Generative AI, and in particular LLMs have gained significant popularity
among researchers and industrial communities, paving the way
for integrating LLMs in different domains, such as robotics~\cite{zeng2023},
telecom~\cite{zou2023}, and healthcare~\cite{javaid2023}.

If we take a step back, we notice that in recent years the rapid advancement of artificial intelligence and machine learning technologies has led to delegating an increasing number of tasks to machines. From automated customer service chatbots to autonomous vehicles, these intelligent systems have demonstrated remarkable capabilities in handling complex tasks and augmenting human productivity. However, as our reliance on those machines grows, it becomes crucial to comprehend the extent of what we are delegating.
LLMs have been employed in various domains, however amidst their impressive capabilities, it is essential to acknowledge the potential drawbacks and limitations they possess.

One of the known drawbacks of LLMs is their capacity to randomize (see for example~\cite{hopkins2023}), therefore we should approach with a certain level of skepticism their performance on tasks when randomization is a critical factor.
In the realm of game theory, randomization plays a pervasive role. A prime illustration of this phenomenon is found in the notion of a mixed strategy Nash equilibrium. This concept entails a strategy wherein a player does not consistently opt for the same action but rather elects each action with a specific probability. This element of uncertainty adds depth and complexity to the strategic decision-making process, as players strategically allocate their choices based on the likelihood of favorable outcomes.

In this work, we focus our attention on the capabilities of LLMs to find the Nash equilibrium in games with a mixed strategy Nash equilibrium and no pure strategy Nash equilibrium (that throughout this work we denote {\sl mixed strategy Nash equilibrium games}). For other works, studying LLMs playing games, see for example~\cite{PlayingGamesWithGPT} and \cite{akata2023}. In particular, we study two classic mixed strategy Nash equilibrium games: matching pennies and rock, paper, scissors, which we describe in the following sections. These simple yet universally recognized games provide an ideal platform to evaluate the LLM capabilities in a controlled environment.

To perform our experiments, we used as LLM the Mistral v0.3 model and the quantized Hermes-2-Pro-Llama-3-8B, an advanced language model that leverages quantization to improve computational efficiency without significantly compromising performance.
We chose this open weights models to ensure that our experiments are designed and conducted in a manner that allows other researchers to replicate them accurately.
This commitment to reproducibility not only validates our findings but also facilitates further research in this area.
Our code can be found on \url{https://github.com/alonsosilvaallende/LLMs_Playing_MSNE_Games}.
Another reason to use these models is that one of the key aspects we use in this paper is the model's function-calling capabilities. This feature allows the model to call functions to answer a user's question. In this manuscript, we only present the results for the Mistral v0.3 model (for Hermes-2-Pro-Llama-3-8B the results are similar and can be found in our code).

\section{Matching Pennies game}

\subsection{Problem description}

Matching Pennies game is a classic game. It is a two-player zero-sum game, meaning that any gain by one player is exactly offset by the loss of the other player. The game works as follows:
\begin{itemize}
    \item[$\bullet$] Each player has two possible actions: to play Heads or Tails.
    \item[$\bullet$] Both players reveal their choices simultaneously.
    \item[$\bullet$] If both choices match (both Heads or both Tails), Player 1 wins. If the choices do not match (one Heads and one Tails), Player 2 wins.
\end{itemize}

The payoff matrix of the Matching Pennies game is given by Table~\ref{tab:MPpayoff}.
\begin{table}
\centering
\caption{Payoff matrix of the Matching Pennies game}
\label{tab:MPpayoff}
\begin{tabular}{|c|c|c|}
\hline
 &  Heads & Tails \\ 
\hline
Heads & (1,-1) & (-1,1) \\ 
\hline
Tails & (-1,1) & (1,-1) \\ 
\hline
\end{tabular}
\end{table}

In the Matching Pennies game, the mixed strategy Nash equilibrium occurs when each player chooses Heads or Tails with equal probability (50\% chance of Heads and 50\% chance of Tails). 
This is because, given that the other player is playing a 50-50 strategy, each player's expected payoff is the same regardless of their own strategy. Therefore, they have no incentive to deviate from the 50-50 strategy, which makes it a Nash equilibrium.

\subsection{Game-playing capabilities of the LLM based purely on its pre-existing knowledge}

In this subsection, we investigate the capabilities of the LLM to play the game. 
First, we assess the LLM's ability to recognize the game to see if it is included in its pre-existing knowledge. To do this we use the following prompt:

\begin{mdframed}
\verb|PROMPT:| What is the mixed strategy Nash equilibrium for the 
matching pennies game?
\end{mdframed}

The response generated by the LLM indicates that indeed the Nash Equilibrium of this game is included in its pre-existing knowledge:
\begin{mdframed}
\verb|ANSWER:| The Matching Pennies game is a two-player, zero-sum game where each player secretly chooses either heads (H) or tails (T). If both players choose the same option, the player who chose first wins the coin, and if they choose differently, the second player wins.

In a Nash equilibrium, neither player has an incentive to deviate from their strategy if they believe the other player will stick to theirs. However, in the Matching Pennies game, there is no pure strategy Nash equilibrium because both players would have an incentive to deviate from their strategies if they knew what the other player was going to do.

Instead, the Matching Pennies game has a mixed strategy Nash equilibrium. In this case, each player should randomly choose heads or tails with equal probability (0.5). This way, neither player can gain an advantage by deviating from their random choice, as they have no information about the other player's decision.

So, the mixed strategy Nash equilibrium for the Matching Pennies game is each player choosing H and T with a 50\% probability.
\end{mdframed}

We assess the model's ability to play the game of matching pennies without any additional instruction. This evaluation is significant as it tests the model's capacity to comprehend and follow the game's rules based purely on its pre-existing knowledge. To do this we use the following prompt:
\begin{mdframed}
\verb|PROMPT:| Respond to the following text with one of the following answers: `heads', `tails'. Provide your response in JSON format containing a single key `answer'. Text: Let's play matching pennies. I have made my choice. Now, you should make your choice. What do you play? 
\end{mdframed}
The idea of using JSON format with a single key is to make the postprocessing of the LLM's answers easier (we just need to extract the answer from the JSON instead of having to search within a potentially long answer which was the action played by the LLM). This idea was taken from the Scikit-LLM library~\cite{sk-llm}.

We tested $1000$ different seeds. The probabilities of obtaining Heads or Tails are given by Table~\ref{tab:LLM_MP}.
\begin{table}
\centering
\caption{Probabilities of obtaining Heads or Tails in the Matching Pennies game without any additional instruction.}
\label{tab:LLM_MP}
\begin{tabular}{|c|c|}
\hline
$p_{\rm Heads}$ & 77.9\% \\
\hline
$p_{\rm Tails}$ & 22.1\% \\
\hline
\end{tabular}
\end{table}

We also tested a different prompt explicitly telling the LLM to play according to the mixed strategy Nash equilibrium of the game.
\begin{mdframed}
\verb|PROMPT:| Respond to the following text with one of the following answers: `heads', `tails'. Provide your response in JSON format containing a single key `answer'. Text: Let's play matching pennies. I have made my choice. Now, you should make your choice. Play according to the mixed strategy Nash equilibrium. What do you play? 
\end{mdframed}
We tested $1000$ different seeds. The probabilities of obtaining Heads or Tails are given by Table~\ref{tab:LLM_MP_additional}.
\begin{table}
\centering
\caption{Probabilities of obtaining Heads or Tails in the Matching Pennies game with explicit prompt.}
\label{tab:LLM_MP_additional}
\begin{tabular}{|c|c|}
\hline
$p_{\rm Heads}$ & 81.6\% \\
\hline
$p_{\rm Tails}$ & 18.4\% \\
\hline
\end{tabular}
\end{table}

The responses generated by the LLM indicate that despite both the Matching Pennies game and its mixed strategy Nash equilibrium being included in its pre-existing knowledge, the LLM is unable to play according to this strategy.

\subsection{Let LLMs run code}

In this subsection, we allow the LLM the possibility to run code. To encourage the LLM to use this possibility we add the following text to the prompt:
\begin{center}
    \verb|Choose your choice first in valid Python code.|
\end{center}

Therefore, we use the following prompt:

\begin{mdframed}
\verb|PROMPT:| Let's play matching pennies. I have made my choice. Now, you should make your choice. Choose your choice first in valid Python code.
\end{mdframed}

The response generated by the LLM is the following:
\begin{mdframed}
\begin{verbatim}
import random
choice = random.randint(0,1)
print("Heads" if choice == 0 else "Tails")
\end{verbatim}
\end{mdframed}
This response indicates that despite the LLM being unable to play according to the mixed strategy Nash equilibrium, it does not have a problem generating the code to play the game according to the mixed strategy Nash equilibrium.
Therefore, if we allow the LLM to run the code it generated, it indeed plays the mixed strategy Nash equilibrium.

\subsection{Modified Matching Pennies game}

In this subsection, we consider a modification of the Matching Pennies (that we call the Modified Matching Pennies game) with the payoff matrix given by Table~\ref{tab:MMPpayoff}.
\begin{table}
\centering
\caption{Payoff matrix of the Modified Matching Pennies game}
\label{tab:MMPpayoff}
\begin{tabular}{|c|c|c|}
\hline
 &  Heads & Tails \\ 
\hline
Heads & (7,-1) & (-1,1) \\ 
\hline
Tails & (-1,1) & (1,-1) \\ 
\hline
\end{tabular}
\end{table}

In this Modified Matching Pennies game, the only modification is that if both players choose Heads, the first player receives a payoff of 7 instead of 1 in the original matching pennies game. All the other payoffs are the same.

The mixed strategy Nash equilibrium of the Modified Matching Pennies~\cite{spaniel2011} is given by:
\begin{itemize}
    \item[$\bullet$] Player 1 plays both options (Heads and Tails) with equal probability~$1/2$.
    \item[$\bullet$] Player 2 plays Heads with probability $1/5$ and Tails with probability $4/5$.
\end{itemize}

To make the LLM play the Modified Matching Pennies Game, we need to be more explicit on the payoff matrix since the LLM does not know it. We also need to add to the prompt the phrase "Choose your choice first in valid Python code." to encourage it to use this possibility. Therefore, we use the following prompt:
\begin{mdframed}
\verb|PROMPT:| 
Let's play a modification of the Matching Pennies game. More precisely, each of us has two possible actions: to play Heads or Tails. We both reveal our choices simultaneously. If both of us play Heads, then I gain 7 points and you lose 1 point, if both of us play Tails, then I gain 1 point and you lose 1 point. If our choices do not match (one Heads and one Tails), you gain 1 point and I lose 1 point. I have made my choice. Now, you should make your choice. What do you play? Choose your choice first in valid Python code.
\end{mdframed}
The response generated by the LLM is the following:
\begin{mdframed}
\begin{verbatim}
import random
choice = "Heads" if random.randint(0,1) == 0 else "Tails"
print("I have played", choice)
\end{verbatim}
\end{mdframed}
This response is the same as the unmodified classic Matching Pennies game. The LLM is unable to find the randomization needed to play the mixed strategy Nash equilibrium of this slightly modified game.

\section{Rock, Paper, Scissors game}

\subsection{Problem description}
The classic game of Rock, Paper, Scissors (RPS) is also a two-player zero-sum game. The game works as follows:
\begin{itemize}
    \item[$\bullet$] Each player has three possible actions: Rock, Paper, or Scissors.
    \item[$\bullet$] Both players reveal their choices simultaneously.
    \item[$\bullet$] Rock crushes Scissors, Scissors cuts Paper, and Paper covers Rock.
\end{itemize}

The payoff matrix of the original Rock, Paper, Scissors game is given by Table~\ref{tab:RPSpayoff}. Given that payoff matrix, it is well-known that the mixed strategy Nash equilibrium for this game is for each player to play with equal probability of~$1/3$ each of the~$3$ possible actions.

\begin{table}
\centering
\caption{Payoff matrix of the Rock, Paper, Scissors game}
\label{tab:RPSpayoff}
\begin{tabular}{|c|c|c|c|}
\hline
 & Rock & Paper & Scissors \\
\hline
Rock & (0, 0) & (-1, 1) & (1, -1) \\
\hline
Paper & (1, -1) & (0, 0) & (-1, 1) \\
\hline
Scissors & (-1, 1) & (1, -1) & (0, 0) \\
\hline
\end{tabular}
\end{table}


\subsection{Game-playing capabilities of the LLM based purely on its pre-existing knowledge}

In this subsection, we investigate the capabilities of the LLM to play the game. 
First, we assess the LLM's ability to recognize the game to see if it is included in its pre-existing knowledge. To do this we use the following prompt:

\begin{mdframed}
\verb|PROMPT:| What's the Mixed Strategy Nash equilibrium for the rock, paper, scissors game?
\end{mdframed}

The response generated by the LLM indicates that indeed the Nash Equilibrium of this game is included in its pre-existing knowledge:
\begin{mdframed}
    \verb|ANSWER:| In a one-shot Rock, Paper, Scissors (RPS) game between two players, there is no pure strategy Nash equilibrium because if both players choose the same strategy, they can be beaten by a deviation to another strategy. However, in repeated games with perfect information and no bounds on the number of repetitions, a mixed strategy Nash equilibrium can exist.

In this case, each player randomly chooses their move (rock, paper, scissors) with probabilities that make it unprofitable for either player to deviate from their chosen strategy. One possible set of such probabilities is:

- Player 1 plays rock with a probability of 1/3, paper with a probability of 1/3, and scissors with a probability of 1/3.
- Player 2 plays rock with a probability of 1/3, paper with a probability of 1/3, and scissors with a probability of 1/3.

In this mixed strategy Nash equilibrium, neither player has an incentive to deviate from their chosen strategy because the expected payoff for each move is the same against every possible move of the other player. This results in a draw on average, as each player wins approximately one-third of the games against the other player's strategies.
\end{mdframed}

We assess the model's ability to play the rock, paper, scissors game without any additional instruction. This evaluation is significant as it tests the model's capacity to comprehend and follow the game's rules based purely on its pre-existing knowledge. To do this we use the following prompt:
\begin{mdframed}
\verb|PROMPT:| Let's play rock, paper, scissors. I have made my choice. Now, you should make your choice. What do you play?
\end{mdframed}
We tested $1000$ different seeds. The probabilities of obtaining rock, paper or scissors are given by Table~\ref{tab:LLM_RPS}.
\begin{table}
\centering
\caption{Probabilities of obtaining Rock, Paper, or Scissors without any additional instruction}
\label{tab:LLM_RPS}
\begin{tabular}{|c|c|}
\hline
$p_{\rm Rock}$ & 51.6\% \\
\hline
$p_{\rm Paper}$ & 0.0\% \\
\hline
$p_{\rm Scissors}$ & 48.4\% \\
\hline
\end{tabular}
\end{table}

We also tested a different prompt explicitly telling the LLM to play according to the mixed strategy Nash equilibrium of the game:
\begin{mdframed}
\verb|PROMPT:| Let's play rock, paper, scissors. I have made my choice. Now, you should make your choice. Play according to the mixed strategy Nash equilibrium. What do you play?
\end{mdframed}
We tested 1000 different seeds. The probabilities of obtaining rock, paper, or scissors are given by Table~\ref{tab:LLM_RPS_additional}.
\begin{table}
\centering
\caption{Probabilities of obtaining Rock, Paper, or Scissors without any additional instruction}
\label{tab:LLM_RPS_additional}
\begin{tabular}{|c|c|}
\hline
$p_{\rm Rock}$ & 6.7\% \\
\hline
$p_{\rm Paper}$ & 18.7\% \\
\hline
$p_{\rm Scissors}$ & 74.6\% \\
\hline
\end{tabular}
\end{table}

The responses generated by the LLM indicate that despite both the matching pennies game and its mixed strategy Nash equilibrium being included in its pre-existing knowledge, the LLM is unable to play according to this strategy.

\subsection{Let LLMs run code}

In this subsection, similar to the Matching Pennies game, we allow the LLM the possibility to run code. To encourage the LLM to use this possibility, we use the following prompt:

\begin{mdframed}
\verb|PROMPT:| Let's play rock, paper, scissors. I have made my choice. Now, you should make your choice. Choose your choice first in code.
\end{mdframed}

The response generated by the LLM is the following:
\begin{mdframed}
\begin{verbatim}
import random
choice = ["rock", "paper", "scissors"]
computer_choice = random.choice(choice)
user_input = input("Enter your choice (rock, paper or scissors): ")
if user_input == computer_choice:
 print("It's a tie!")
elif (user_input == "rock" and computer_choice == "scissors") or \
 (user_input == "paper" and computer_choice == "rock") or \
 (user_input == "scissors" and computer_choice == "paper"):
 print("You win!")
else:
 print("Computer wins!")
\end{verbatim}
    
\end{mdframed}

Similar to the case of the Matching Pennies game, this response indicates that despite the LLM being unable to play according to the mixed strategy Nash equilibrium, it does not have a problem generating the code to play the game according to the mixed strategy Nash equilibrium.

\subsection{Modified Rock, Paper, Scissors Game}

In this subsection, we consider a modification of the Rock, Paper, Scissors game (that we call the Modified Rock, Paper, Scissors game) with the payoff matrix given by Table~\ref{tab:MRPSpayoff}.
\begin{table}
\centering
\caption{Payoff matrix of the Modified Rock, Paper, Scissors game}
\label{tab:MRPSpayoff}
\begin{tabular}{|c|c|c|c|}
\hline
 & Rock & Paper & Scissors \\
\hline
Rock & (0, 0) & (-2, 2) & (1, -1) \\
\hline
Paper & (2, -2) & (0, 0) & (-1, 1) \\
\hline
Scissors & (-1, 1) & (1, -1) & (0, 0) \\
\hline
\end{tabular}
\end{table}

To make the LLM play the Modified Rock, Paper, Scissors Game, we need to be more explicit on the payoff matrix since the LLM does not know it. We also need to add to the prompt the phrase "Choose your choice first in valid Python code." to encourage it to use this possibility. Therefore, we use the following prompt:

\begin{mdframed}
\verb|PROMPT:| 
Let's play once a modification of rock, paper, scissors. More precisely, if I play rock and you play scissors, then I gain 1 dollar and you lose 1 dollar, if I play scissors and you play paper, then I gain 1 dollar and you lose 1 dollar, however, if I play paper and you play rock then I gain 2 dollars and you lose 2 dollars. The same holds by changing our roles. I have made my choice. Now, you should make your choice. What do you play? Choose your choice first in valid Python code.
\end{mdframed}

The mixed strategy Nash equilibrium of the Modified Rock, Paper, Scissors game~\cite{spaniel2011} is given by Table~\ref{tab:NE_MRPS}.
\begin{table}
\centering
\caption{Nash equilibrium strategy for the Modified Rock, Paper, Scissors game with payoff matrix given by Table~\ref{tab:MRPSpayoff}}
\label{tab:NE_MRPS}
\begin{tabular}{|c|c|}
\hline
$p_{\rm Rock}$ & 1/4 \\
\hline
$p_{\rm Paper}$ & 1/4 \\
\hline
$p_{\rm Scissors}$ & 1/2 \\
\hline
\end{tabular}
\end{table}

The response generated by the LLM is the following:
\begin{mdframed}
\begin{verbatim}
import random
choice = ['rock', 'paper', 'scissors']
user_choice = input('Enter your choice (rock, paper or scissors): ')
computer_choice = choice[random.randint(0, 2)]
if user_choice == computer_choice:
 print('It is a tie!')
elif (user_choice == 'rock' and computer_choice == 'scissors') or \
 (user_choice == 'scissors' and computer_choice == 'paper') or \
  (user_choice == 'paper' and computer_choice == 'rock'):
 print('You lose 1 dollar. The computer plays', computer_choice)
else:
 print('You win 1 dollar. The computer plays', computer_choice)
\end{verbatim}
\end{mdframed}

This response is similar to the response of the unmodified classic Rock, Paper, Scissors game. The LLM is unable to find the randomization needed to play the mixed strategy Nash equilibrium of this slightly modified game (nor the modification of the payments).

\section{Conclusions}

Our study on the intersection of game theory and generative artificial intelligence, particularly focusing on Large Language Models (LLMs), has provided valuable insights into the capabilities and limitations of LLMs in identifying Nash equilibria in mixed strategy games. Our findings demonstrate that LLMs can significantly enhance their performance when they are enabled to run code and are given specific prompts that encourage this functionality. This capability allows LLMs to perform well in standard game scenarios where the strategies are well-defined and well-documented.

However, the study also highlights critical limitations in the adaptability of LLMs when confronted with games that involve complex randomization strategies or slight modifications from standard scenarios. In such cases, the performance of LLMs noticeably declines, suggesting that while LLMs are proficient in handling familiar and straightforward game dynamics, their effectiveness is reduced in more complex or altered game setups.

This research underscores the necessity for ongoing development in the field of generative AI to enhance the robustness and flexibility of LLMs. Future research should focus on improving the ability of LLMs to handle a broader array of game types, particularly those that deviate from standard forms, to fully leverage the potential of LLMs in practical and theoretical applications. Additionally, further studies are required to explore the integration of advanced machine learning techniques that could aid LLMs in better understanding and adapting to complex game strategies.

Overall, our work contributes to the expanding knowledge base at the intersection of game theory and artificial intelligence and opens up new avenues for research in enhancing the capabilities of generative AI systems in complex decision-making scenarios.

\begin{credits}
\subsubsection{\ackname} The work presented in this paper has been partially carried out at LINCS (\href{http://www.lincs.fr}{http://www.lincs.fr}).
\end{credits}
%
%
%

%

%
%

\bibliographystyle{splncs04}
\bibliography{mybibliography}

%





\end{document}